\documentclass[a4paper,11pt]{article}
\pdfoutput=1 

\usepackage{jinstpub} 
\usepackage{graphbox}

\title{\boldmath The Monopix chips: Depleted monolithic active pixel sensors with a column-drain read-out architecture for the ATLAS Inner Tracker upgrade}

\author[a,1]{I. Caicedo,\note{Corresponding author.}}
\author[b]{M. Barbero,}
\author[b]{P. Barrillon,}
\author[d]{I. Berdalovic,}
\author[b]{S. Bhat,}
\author[a]{C. Bespin,}
\author[b]{P. Breugnon,}
\author[d]{R. Cardella,}
\author[b]{Z. Chen,}
\author[c]{Y. Degerli,}
\author[a]{J. Dingfelder,}
\author[b]{S. Godiot,}
\author[c]{F. Guilloux,}
\author[a]{T. Hirono,}
\author[a]{T. Hemperek,}
\author[a]{F. H\"{u}gging,}
\author[a]{H. Kr\"{u}ger,}
\author[d]{T. Kugathasan,}
\author[a]{K. Moustakas,}
\author[b]{P. Pangaud,}
\author[d]{H. Pernegger,}
\author[a]{D.-L. Pohl,}
\author[d]{P. Riedler,}
\author[b]{A. Rozanov,}
\author[a]{P. Rymaszewski,}
\author[c]{P. Schwemling,}
\author[d]{W. Snoeys,}
\author[c]{M. Vandenbroucke,}
\author[a]{T. Wang,}
\author[a]{N. Wermes}

\affiliation[a]{Institute of Physics, University of Bonn,\\Nussallee 12, Bonn, Germany}
\affiliation[b]{Aix Marseille University, CNRS/IN2P3, CPPM,\\163 Avenue de Luminy, Marseille, France}
\affiliation[c]{IRFU CEA-Saclay,\\Batiment 141, Gif-sur-Yvette Cedex, France}
\affiliation[d]{CERN Experimental Physics Department,\\Espl. des Particules 1, Geneve, Switzerland}

\emailAdd{caicedo@physik.uni-bonn.de}

\abstract{Two different depleted monolithic CMOS active pixel sensor (DMAPS) prototypes with a fully synchronous column-drain read-out architecture were designed and tested: LF-Monopix and TJ-Monopix. These chips are part of a R\&D effort towards a suitable implementation of a CMOS DMAPS for the HL-LHC ATLAS Inner Tracker. LF-Monopix was developed using a 150nm CMOS process on a highly resistive substrate (>2 k$\Omega\,$cm), while TJ-Monopix was fabricated using a modified 180 nm CMOS process with a 1 k$\Omega\,$cm epi-layer for depletion. The chips differ in their front-end design, biasing scheme, pixel pitch, dimensions of the collecting electrode relative to the pixel size (large and small electrode design, respectively) and the placement of read-out electronics within such electrode. 
	
Both chips were operational after thinning down to 100 $\mathrm{\mu}$m and additional back-side processing in LF-Monopix for total bulk depletion. The results in this work include measurements of their leakage current, noise, threshold dispersion, response to minimum ionizing particles and efficiency in test beam campaigns. In addition, the outcome from measurements after irradiation with neutrons up to a dose of $1\times10^{15}\,\mathrm{n_{eq} / cm}^{2}$  and its implications for future designs are discussed.}


\proceeding{PIXEL 2018 International Workshop\\
December 10-14, 2018\\
Activity Center of the Academia Sinica, Taipei, Taiwan}

\begin{document}
\maketitle
\flushbottom

\section{Introduction}
\label{sec:intro}

The foreseen high-luminosity upgrade of the ATLAS experiment at the Large Hadron Collider requires improvements in terms of production costs and radiation hardness of the inner tracker detector (ITk) systems~\cite{atlas_upgrade2}. The area of the pixel modules will be increased and the instantaneous luminosity is expected to reach values about ten times its current nominal value. For the ATLAS ITk, from its outer to innermost layers, these conditions will translate into a particle occupancy from 1 to 30 $\mathrm{MHz/mm^{2}}$, non-ionising displacement damage (NIEL) between $10^{15}$ and $10^{16} \,\mathrm{n_{eq} / cm}^{2}$ and doses from ionizing radiation (TID) between $80$ and $10^{3}$ Mrad by the end of its lifetime.

Ongoing R\&D efforts aim to develop depleted monolithic active pixel detectors (DMAPS) in commercial CMOS processes that could cope with the operation requirements at the outermost layer of the ATLAS ITk. In these devices, the possibility to apply large voltages in highly resistive substrates enables a fast charge collection by drift and improves their radiation hardness. Moreover, the use of nested wells to place the sensor and read-out electronics in a common silicon bulk is attractive in terms of yield and tracking capability, as it avoids the hybridization stage and reduces the material budget.

The Monopix chips are large scale DMAPS prototypes with a fully synchronous column-drain read-out architecture~\cite{columndrain_mandelli}. The conservative but reliable implementation of such architecture in these devices was motivated by preliminary simulations and the successful operation of the FE-I3 chip~\cite{fei3_peric} read-out at similar hit rates. In both chips, when a pixel fires upon detection of a particle hit, the value of a 40 MHz timestamp distributed over the pixel matrix is registered both on the rising and falling edge of the comparator output. The difference between these two values represents the Time-Over-Threshold (ToT) and it can be used for a coarse measurement of the analog response. 

The key feature defining the front-end specifications, advantages and limitations of each Monopix design is the size of the collecting electrode within the pixels. This value influences each pixel's detector capacitance, which in turn is proportional to its noise and shaping time degradation. On one side, a large electrode layout demands power as a trade-off for the increase in detector capacitance but improves field uniformity. On the other, a small electrode design increases the drift path and trapping probability for charge collection, compromising its radiation-hardness. TJ- and LF- Monopix have different pixel layouts, front-end implementations and biasing schemes (labeled as "flavours") which are optimized to compensate the limitations in each approach.

\subsection{LF-Monopix}

LF-Monopix is the first large scale DMAPS prototype with fully integrated electronics for fast standalone readout in its pixel units~\cite{lfmono_rymaszewski}. It was fabricated in a 150 nm CMOS process from LFoundry on a high-resistivity (> 2  k$\Omega\,$cm) p-type substrate and its guard ring layout was optimized to allow bias voltages larger than 250 V for depletion. The radiation hardness of the process after TID and NIEL damage was demonstrated and optimized in previous prototypes~\cite{ccpdlf_hironofirst, lfmono_hirono}.

\begin{figure}[htbp]
	\centering
	\small		
		\includegraphics[width=0.64\linewidth]{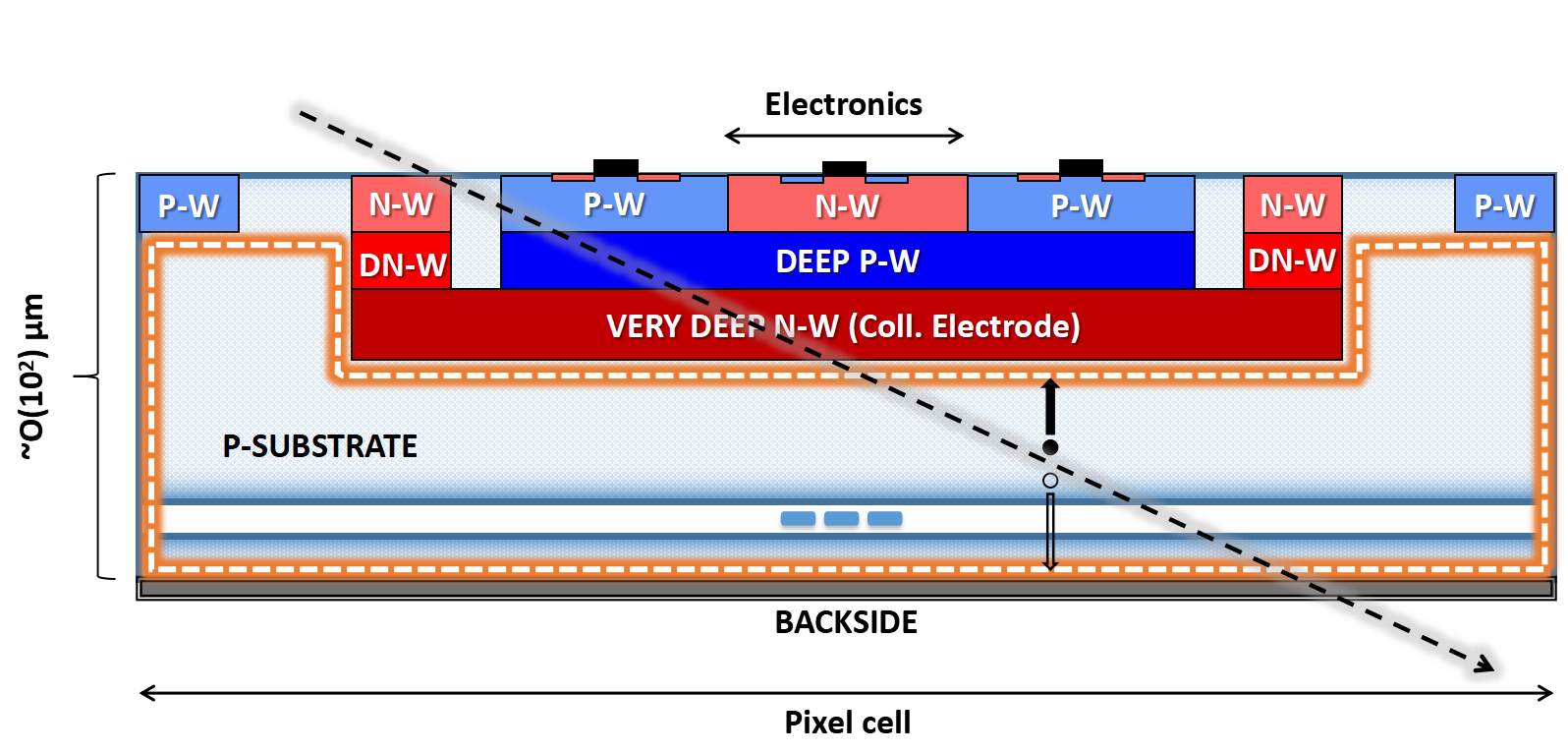}
	\caption{\label{fig:lfmonopixlayout} Pixel cross-section of LF-Monopix. The depletable volume is delimited by white dashed lines.}
\end{figure}

The chip matrix consists of an array of $129\times36$ pixels with an individual pitch of $\mathrm{250\times50\,\mu m^{2}}$. The design of each cell follows a large electrode approach, as depicted in Figure~\ref{fig:lfmonopixlayout}: Charge is created in a depleted p-type substrate and it drifts towards a large n-type collection electrode. The integrated read-out electronics are located inside the electrode while isolated from it by p-type wells in order to avoid signal coupling. 

The front-end electronics were optimized to cope with a detector capacitance of about $\mathrm{400\,fF}$ and to operate with an average power consumption of $\mathrm{\sim 36\,\mu W/pixel}$. The design showed noise levels around $\mathrm{200\,e-}$ and a gain of $\mathrm{10-12\,\mu V/e-}$ depending on the flavour implementation. The pixel threshold distribution was tunable via a 4-bit local DAC down to $\mathrm{1400\pm100\,e-}$ before irradiation and $\mathrm{1700\pm130\,e-}$ after NIEL damage up to $1\times10^{15} \,\mathrm{n_{eq} / cm}^{2}$.

\subsection{TJ-Monopix}

TJ-Monopix is a DMAPS prototype with pixels using a small electrode for charge collection~\cite{lftjmono_wang, tjmono_moustakas}. The chip was designed in a modified 180nm CMOS sensor process from Towerjazz, where a 1 k$\Omega\,$cm epi-layer can be fully depleted through the addition of a low dose n-type implant~\cite{tjmono_snoeys}. The radiation tolerance of the process was validated through uniform high efficiency measurements on a test chip after neutron irradiation up to a dose of $1\times10^{15}\, \mathrm{n_{eq} / cm}^{2}$~\cite{tjmono_pernegger}.

\begin{figure}[htbp]
	\centering
	\small		
	\begin{tabular}{lc}
		\textbf{RDPW}&
		\includegraphics[align=c,width=0.67\linewidth]{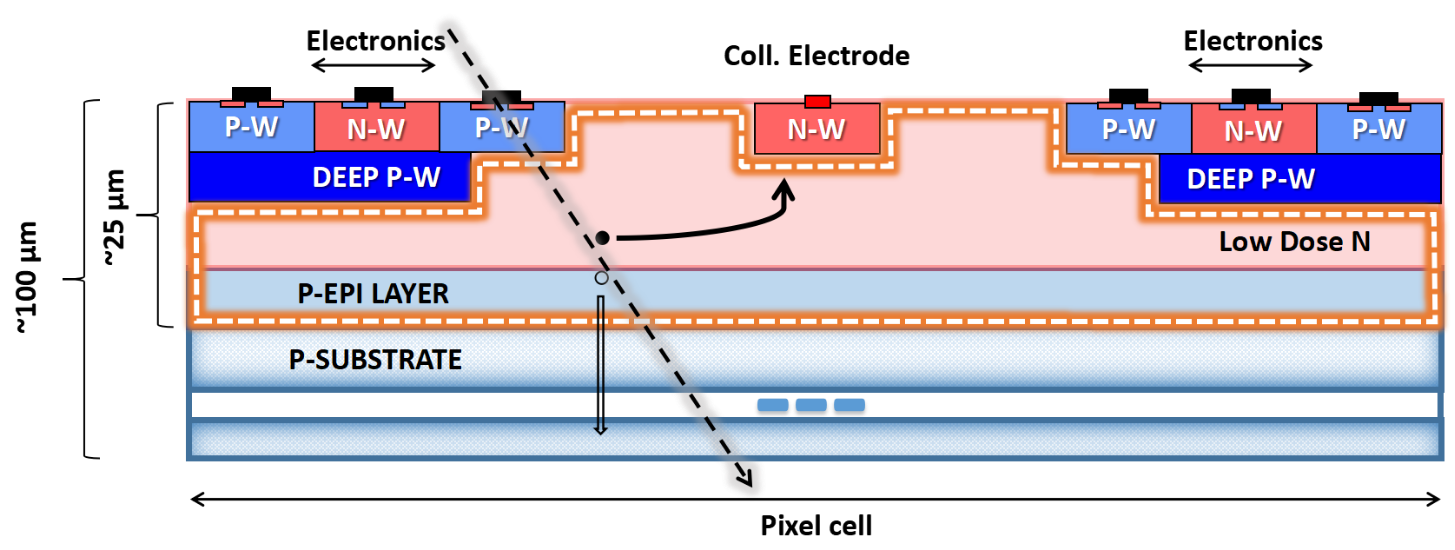}
		\\
		\textbf{FDPW}& 
		\includegraphics[align=c,width=0.67\linewidth]{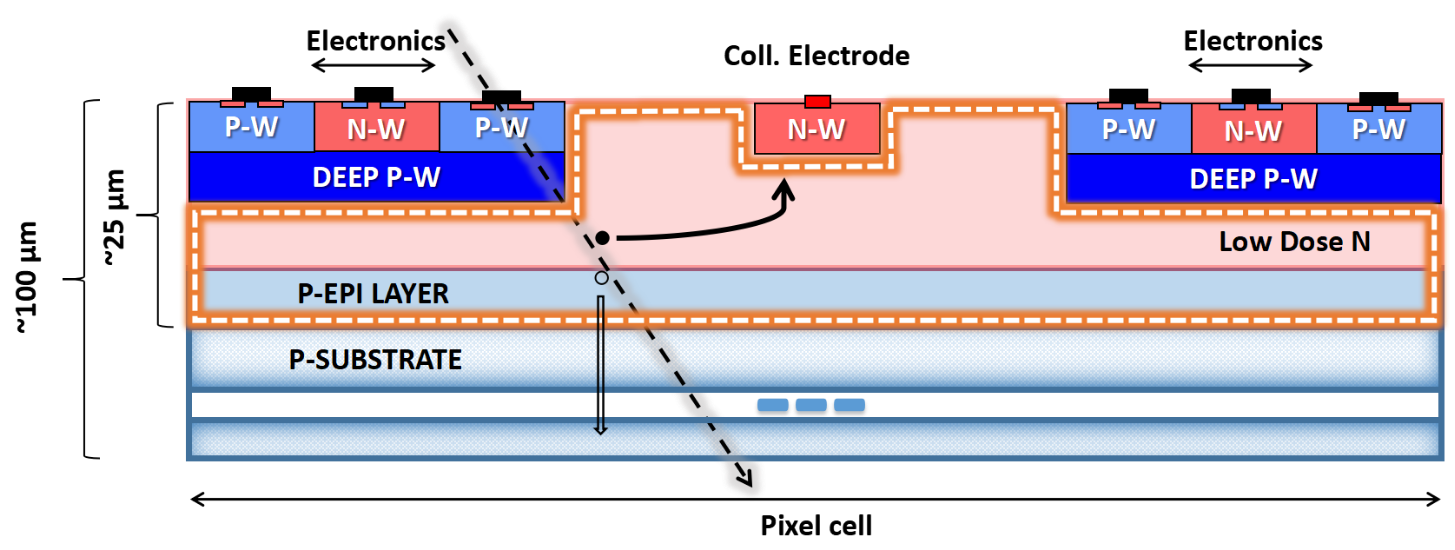}
	\end{tabular}
	\caption{\label{fig:tjmonopixlayout} Pixel cross-sections for the removed deep p-well (RDPW) and full deep p-well (FDPW) layouts in TJ-Monopix. The depletable volumes are delimited by white dashed lines.}
\end{figure}

Figure~\ref{fig:tjmonopixlayout} illustrates the cross-sections and layouts for two different pixel designs in TJ-Monopix. Read-out electronics are separated from the collection node and  isolated from the depletable volume through p-type wells. Pixels in the top half of the chip had a considerable fraction of the deep p-well coverage removed (RDPW) as a test to increase the depleted area and improve the lateral electric field near the collecting node. The bottom half preserved the whole full deep p-well (FDPW) coverage below the read-out electronics as a reference to assess the effect of its removal. 

TJ-Monopix comprises an array of $224\times448$ pixels with a pitch of $\mathrm{36\times40 \,\mu m^{2}}$. It uses a compact low-power front-end  ($\mathrm{\sim 1\,\mu W/pixel}$ from design) optimized to cope with the timing requirements of ATLAS. During tests, the chip showed a gain of $\mathrm{\sim400\,\mu V/e-}$.

\section{Instrumentation and Methods}

The chips were wire-bonded to dedicated single chip cards, which in turn were coupled to a General Purpose Analog Card (GPAC) and Multiple Input/Output (MIO) board. The GPAC provides an injection pulse generator, plus 12-bit DACs, ADCs, power sources and the possibility to probe digital and analog signals. The MIO holds a programmable Kintex7 FPGA, connectors for external triggers and it allows to communicate via Ethernet with a PC. 

Samples from both chips were irradiated with neutrons up to different fluences at the nuclear reactor of the Jozef Stefan Institute in Lubljana. In order to determine their hit detection efficiency before or after irradiation, both sensors were placed within the path of Minimum Ionizing Particle (MIP) beams traversing an EUDAQ-type~\cite{eudet} reference telescope array. Two types of beams were used according to availability within the characterization period: $\mathrm{2.5\,GeV}$ electrons at the ELSA accelerator in Bonn~\cite{ELSA} or $\mathrm{180\,GeV}$ pions in the CERN SPS facility~\cite{SPS}. The minimum spatial resolution achievable for tracks in the devices under test was limited by the telescope resolution ($\sim 5 \,\mu$m) and amount of particle scattering, which is inversely related to the beam energy. In the case of irradiated samples, the chips were placed inside of a sealed styrofoam box capable of reaching down to $\mathrm{-30\,^{\circ} C}$. The analysis of test beam data was carried out using a python-based analysis framework developed in Bonn and thoroughly tested for this type of telescope~\cite{TBA_Bonn}.

\section{Measurements with LF-Monopix}

\subsection{Leakage and depletion after thinning}

\begin{figure}[htbp]
\centering 
\includegraphics[width=.56\textwidth]{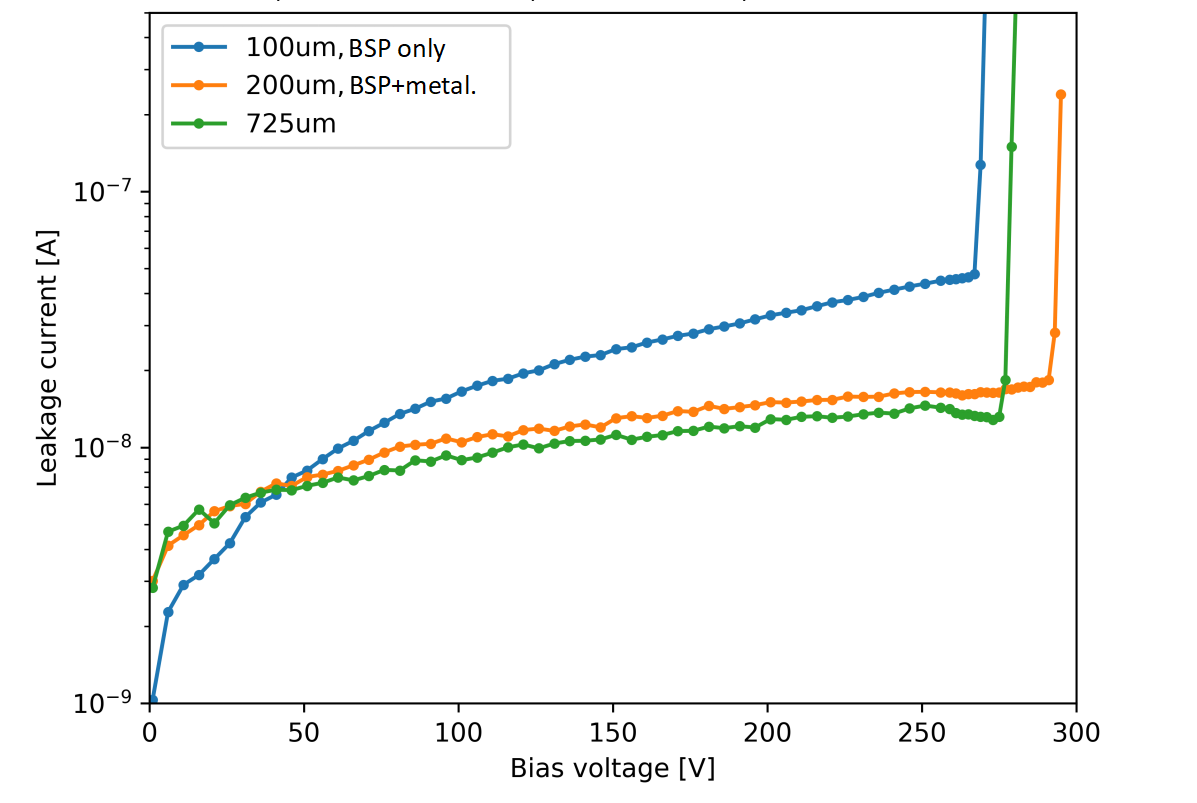}
\caption{\label{fig:lfmonopixIVcurves} I-V curves for back-side processed LF-Monopix samples of different thicknesses: 725$\mathrm{\,\mu}$m, 200$\mathrm{\,\mu}$m (with metallization) and 100$\mathrm{\,\mu}$m (without metallization). All measurements at room temperature.}
\end{figure}

LF-Monopix wafers were succesfully thinned down to 200 or $\mathrm{100\,\mu}$m and back-side processed, with additional metallization in the $\mathrm{200\,\mu}$m ones. As the thickness of the substrate is reduced, the magnitude of the electric field within it is expected to increase. Moreover, previous e-TCT studies on test structures in LFoundry 150 nm CMOS~\cite{lfmono_mandic} suggested that wafer thinning and back-side processing would result in an improvement in charge collection for irradiated thinned samples. By looking at their I-V curves (as depicted in Figure~\ref{fig:lfmonopixIVcurves}), the breakdown voltage was larger than $\mathrm{260\,V}$ independent of the sensor thickness or addition of metallization, in agreement with previous observations for unthinned samples~\cite{lfmono_iguaz, lftjmono_wang}. 

\begin{figure}[htbp]
\centering
\small		
\begin{tabular}{cc}
\includegraphics[width=0.48\linewidth]{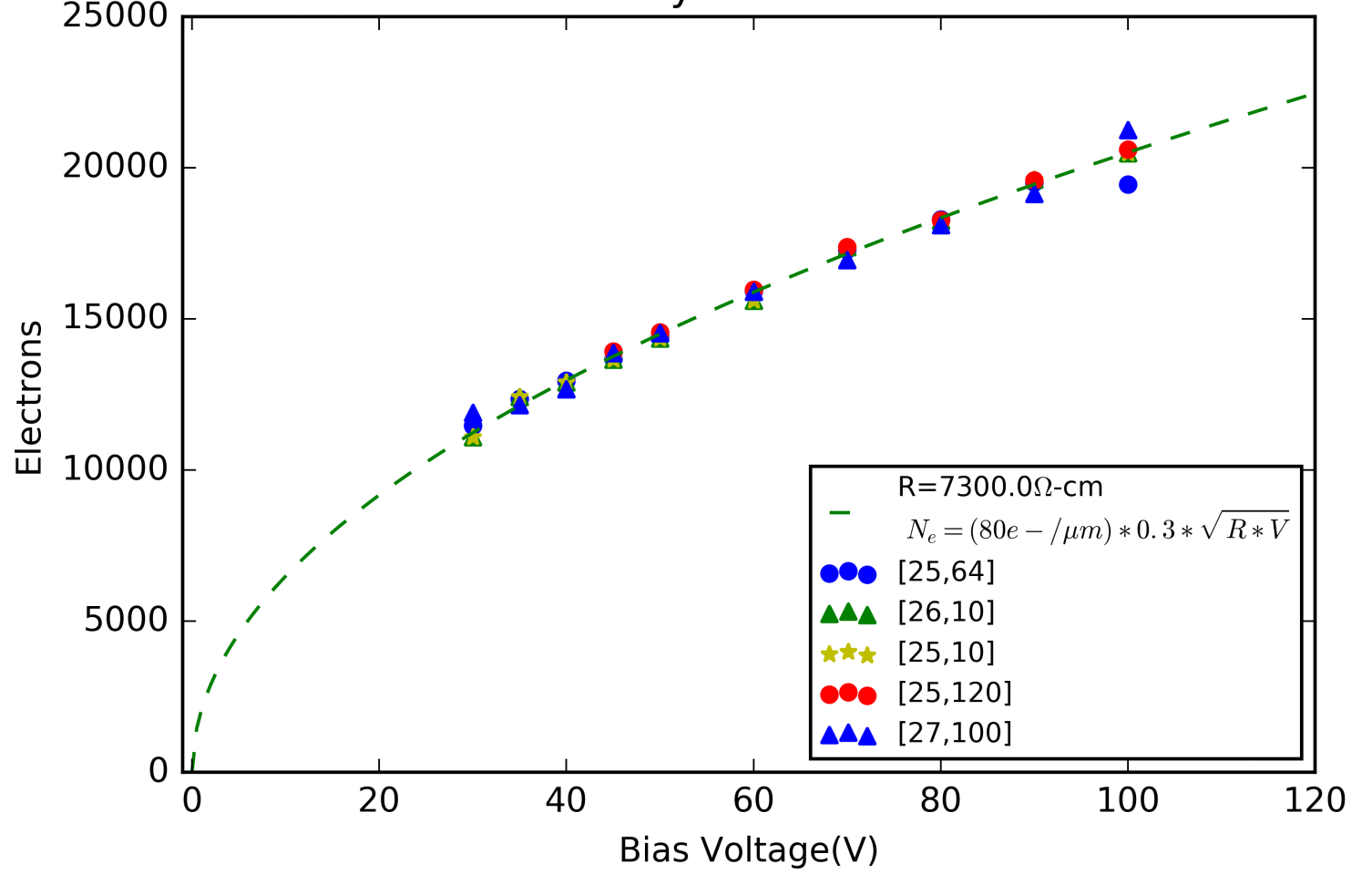}&
\includegraphics[width=0.475\linewidth]{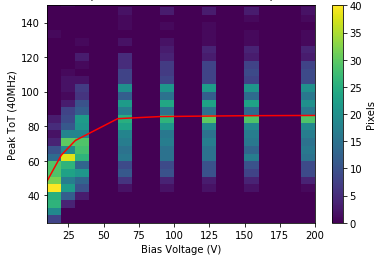}
\\ (a) MPV (in e-) of MIPs in a 725$\mathrm{\,\mu}$m thick chip & (b) MPV (in ToT u.) of MIPs in a 200$\mathrm{\,\mu}$m thick chip\\
\end{tabular}
\caption{\label{fig:lfmonopixdepletion} Depletion measured through the Most Probable Values of MIPs crossing the substrate of LF-Monopix.}
\end{figure}

The depletion properties of LF-Monopix were estimated indirectly by looking at the most probable value (MPV) of the energy deposited by MIPs interacting with the silicon bulk for different bias voltages. The left side of Figure~\ref{fig:lfmonopixdepletion} shows the calibrated MPV of the energy deposited in 5 different pixels by MIPs interacting with a 725$\mathrm{\,\mu}$m thick sample. A fit to these data points resulted in a measured resistivity of 7.3 k$\Omega\,$cm,  which is larger than previously reported values of 3-5 k$\Omega\,$cm (in~\cite{ccpdlf_hirono},~\cite{lfpassive_pohl} and~\cite{lfmono_vigani}) but still in agreement with the claim of a resistivity larger than 2 k$\Omega\,$cm by the foundry. The right side of Figure~\ref{fig:lfmonopixdepletion} shows a distribution of uncalibrated MPVs measured by all pixels in a tuned flavour ( $\mathrm{\sim100\,e-}$ threshold dispersion) of a $\mathrm{200\,\mu}$m thick sample. The saturation of the measured MPVs for voltages equal or larger than 60V comes into good agreement with the independent measurement of $\sim 200\,\mathrm{\mu m}$ of silicon depleted for that voltage in  the 725$\mu m$ thick chip. Based on that result, $100\,\mathrm{\mu m}$ of silicon would be fully depleted with a voltage $\mathrm{\sim15\,V}$, which in turn would suggest that the increase in leakage current of the $\mathrm{100\,\mu}$m thick chip in Figure~\ref{fig:lfmonopixIVcurves} occurred when the sensor without metallization was fully depleted.

\subsection{In-Pixel efficiency in non-irradiated chips}

\begin{figure}[htbp]
\centering
\small		
\includegraphics[width=0.83\linewidth]{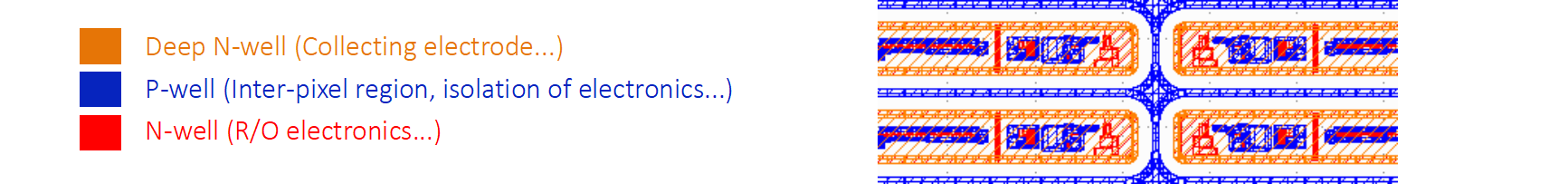}
\\ \textbf{(a)} Pixel layout: Section of a 2x2 pixel array. \\
\begin{tabular}{cc}
\includegraphics[width=0.445\linewidth]{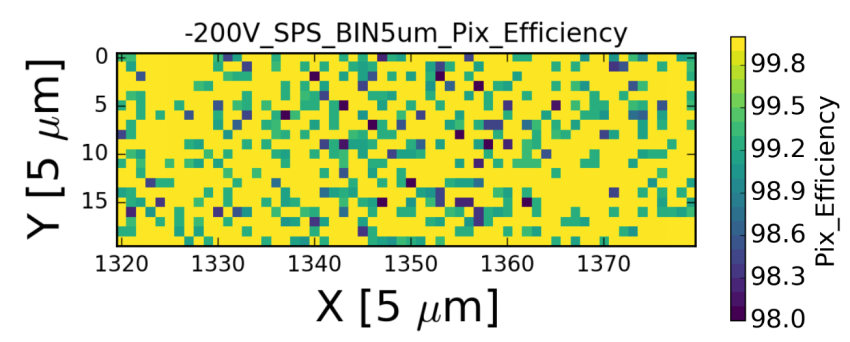}&
\includegraphics[width=0.445\linewidth]{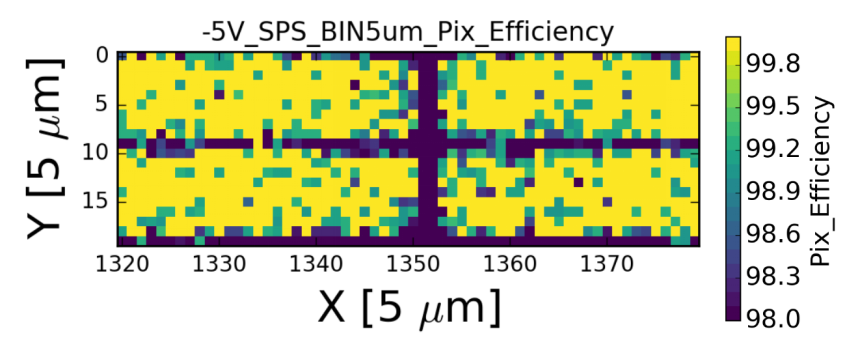}
\\ \textbf{(b)} In-pixel efficiency at -200 V: 99.7 \%  & \textbf{(c)} In-pixel efficiency at -5 V: 98.8 \%\\
\end{tabular}
\caption{\label{fig:lfinpixefficiency} In-pixel efficiency for a non-irradiated LF-Monopix (Threshold: $\mathrm{2700\,e-}$).}
\end{figure}

Mean detection efficiencies for LF-Monopix were already reported for a threshold of $1700\mathrm{\,e-}$~\cite{lfmono_hirono}: $99.6\%$ before irradiation and $98.9\%$ after neutron irradiation with a fluence of $1\times10^{15} \mathrm{\,n_{eq} / cm}^{2}$. The response of a non-irradiated chip to a $\mathrm{180\,GeV}$ pion beam was mapped with sub-pixel spatial resolution (5$\mathrm{\,\mu}$m) for a closer look into the uniformity of the hit detection efficiency across the pixel area, as shown in Figure~\ref{fig:lfinpixefficiency}. Plot ~\ref{fig:lfinpixefficiency}\textbf{(a)} illustrates the pixel layout of the mapped region. Plot~\ref{fig:lfinpixefficiency}\textbf{(b)} shows that for an unirradiated chip with high applied bias voltage the detection efficiency is overall uniform across the pixel surface. Finally, plot~\ref{fig:lfinpixefficiency}\textbf{(c)} shows that localized inefficiencies appear only at the pixel borders when very low bias voltages are applied. These localized losses were also observed with limited resolution in neutron irradiated samples. In both cases, small charge signals created near the edges are shared between neighboring pixels and the probability of a collected signal below threshold -and therefore not registered as a hit- increases.   

\section{Measurements with TJ-Monopix}

\subsection{Threshold and noise}

\begin{figure}[htbp]
	\centering
	\small		
	\begin{tabular}{cc}
		\includegraphics[width=0.45\linewidth]{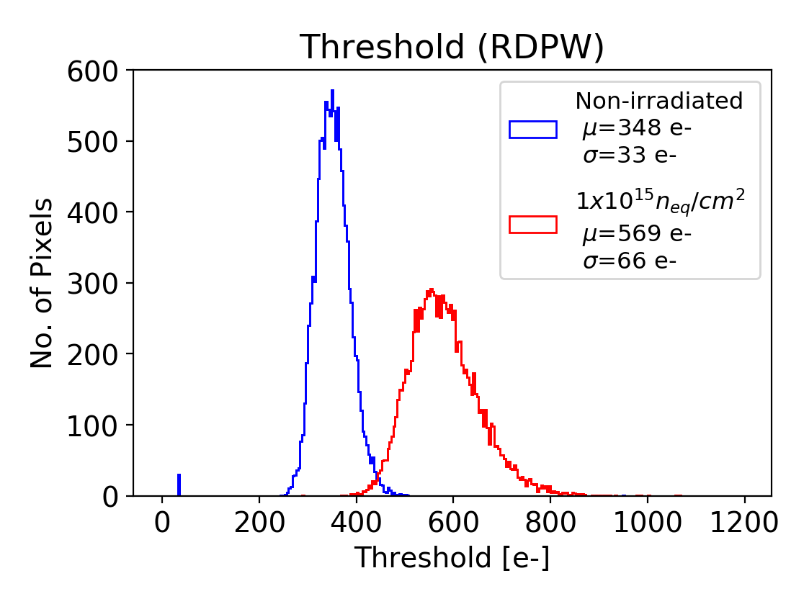}&
		\includegraphics[width=0.45\linewidth]{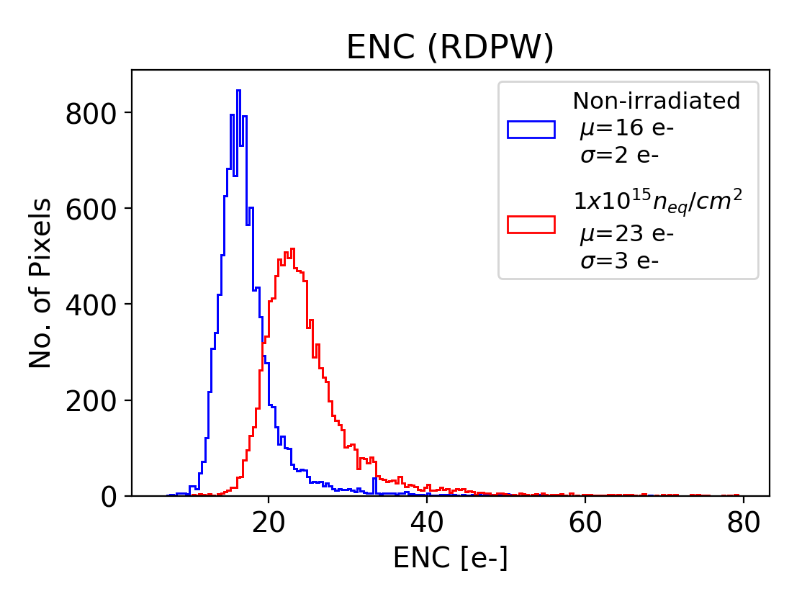}
	\end{tabular}
	\caption{\label{fig:tjTHandENC} Threshold (Left) and ENC (Right) distributions in TJ-Monopix before and after neutron irradiation up to a fluence of $1\times10^{15} \mathrm{\,n_{eq} / cm}^{2}$}
\end{figure}

By means of a scan over injection DAC steps, it was possible to determine the threshold and equivalent noise charge (ENC) in TJ-Monopix from the mean and standard deviation of the resulting s-curves, respectively. The threshold and noise distributions for the RDPW region are shown in Figure ~\ref{fig:tjTHandENC}. The threshold before irradiation had a mean around $\mathrm{350\,e-}$ with a dispersion of $\mathrm{35\,e-}$. This mean increased to $\mathrm{570\,e-}$ after NIEL irradiation up to $1\times10^{15} \mathrm{n_{eq} / cm}^{2}$ and so did its dispersion up to $\mathrm{65\,e-}$. The noise of the non-irradiated chip was in the order of $\mathrm{15\,e-}$, with an increase of $\mathrm{10\,e-}$ after irradiation. These increments are a consequence of the TID background at the reactor ($\mathrm{\sim1MRad}$ for the achieved fluence). Both threshold and noise values were independent of the deep p-well coverage of the chip for the measured implementation.

\subsection{Mean detection efficiency after NIEL irradiation}

\begin{figure}[htbp]
	\centering
	\small		
	\begin{tabular}{cc}
		\includegraphics[width=0.38\linewidth]{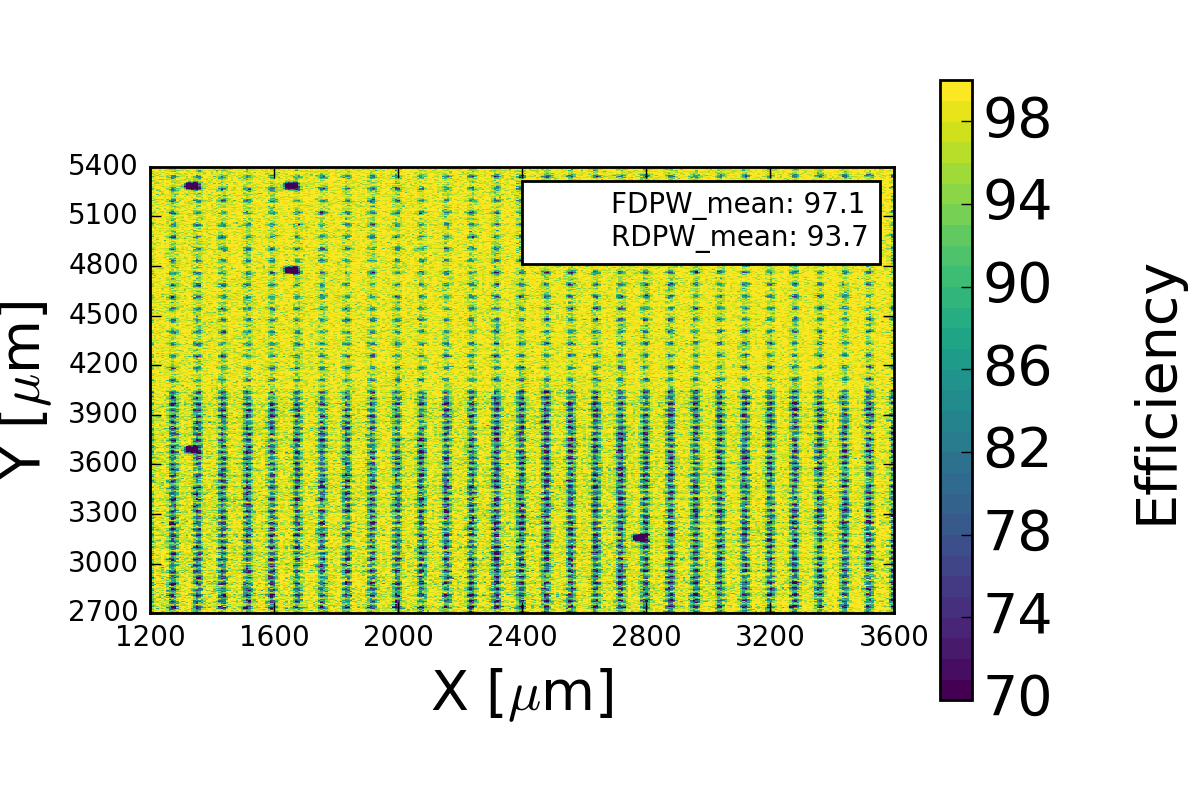}&
		\includegraphics[width=0.38\linewidth]{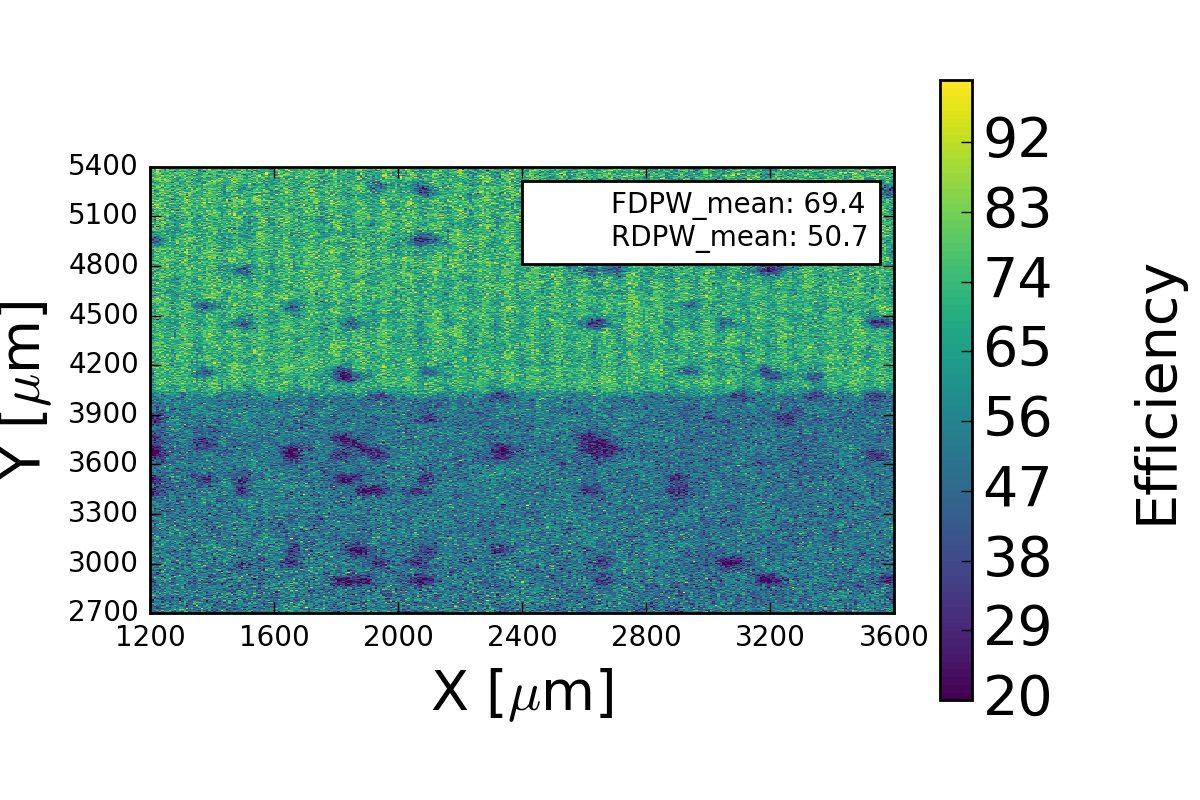}
		\\ \textbf{(a) Non-irradiated} &  \textbf{(b)} \boldmath$1\times10^{15} \mathrm{\,n_{eq} / cm}^{2}$ \\ RDPW: 97.1\% | FDPW: 93.7\%  & RDPW: 69.4\% | FDPW: 50.7\%\\
	\end{tabular}
	\caption{\label{fig:tjIRRefficiency} Mean detection efficiencies before and after neutron irradiation in TJ-Monopix.}
\end{figure}

The mean detection efficiency in TJ-Monopix was calculated out of data obtained at the $\mathrm{2.5\,GeV}$ electron beam in ELSA. Efficiencies before and after NIEL irradiation up to $1\times10^{15} \mathrm{\,n_{eq} / cm}^{2}$ are shown in Figure~\ref{fig:tjIRRefficiency}. It was not possible to map the efficiency of the irradiated sample with sub-pixel resolution due to additional electron scattering in the cooling box. For the non-irradiated chip, the mean detection efficiency was $97.1\%$ in the RDPW region and $93.7\%$ in the FDPW one. After irradiation, the mean efficiency dropped by $\sim$28$\%$ in the RDPW part and $\sim$43$\%$ in the FDPW one.

\subsection{In-Pixel efficiency in non-irradiated chips}

\begin{figure}[htbp]
	\centering
	\small		
	\begin{tabular}{lcc}
		\begin{tabular}{@{}c} \textbf{RDPW}\\(Mean eff: 97.1\%) \end{tabular}&
		\includegraphics[align=c,width=0.24\linewidth]{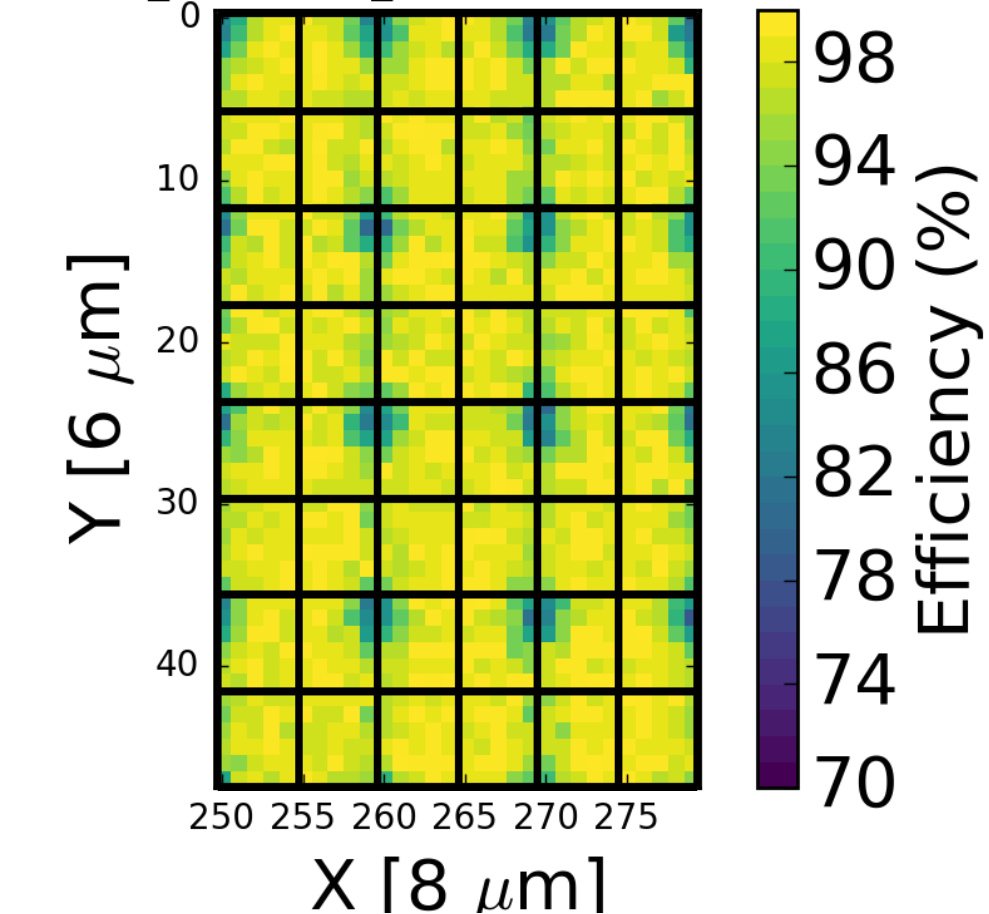}&
		\includegraphics[align=c,width=0.22\linewidth]{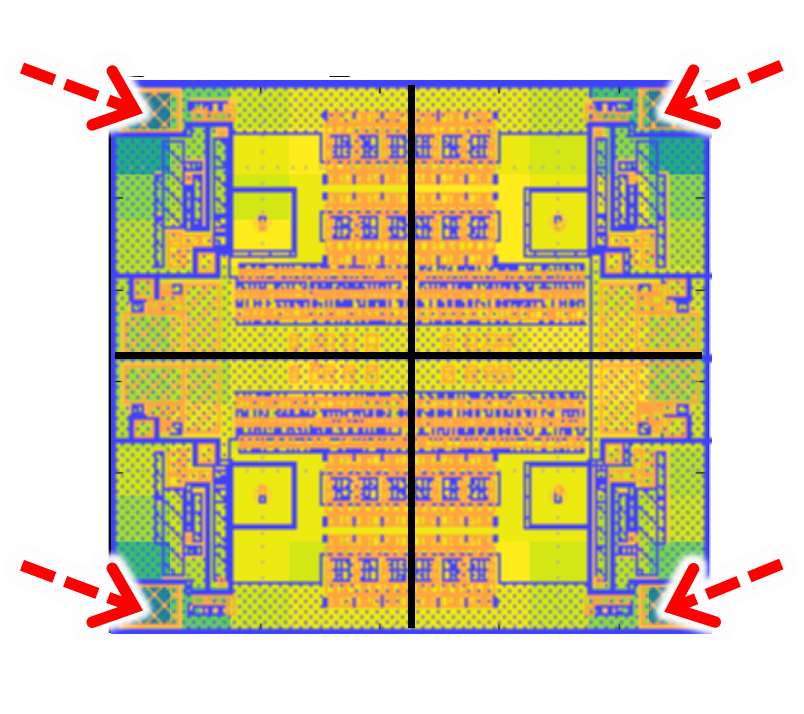}
		\\
		\\
		\begin{tabular}{@{}c} \textbf{FDPW}\\(Mean eff: 93.7\%) \end{tabular}& 
		\includegraphics[align=c,width=0.24\linewidth]{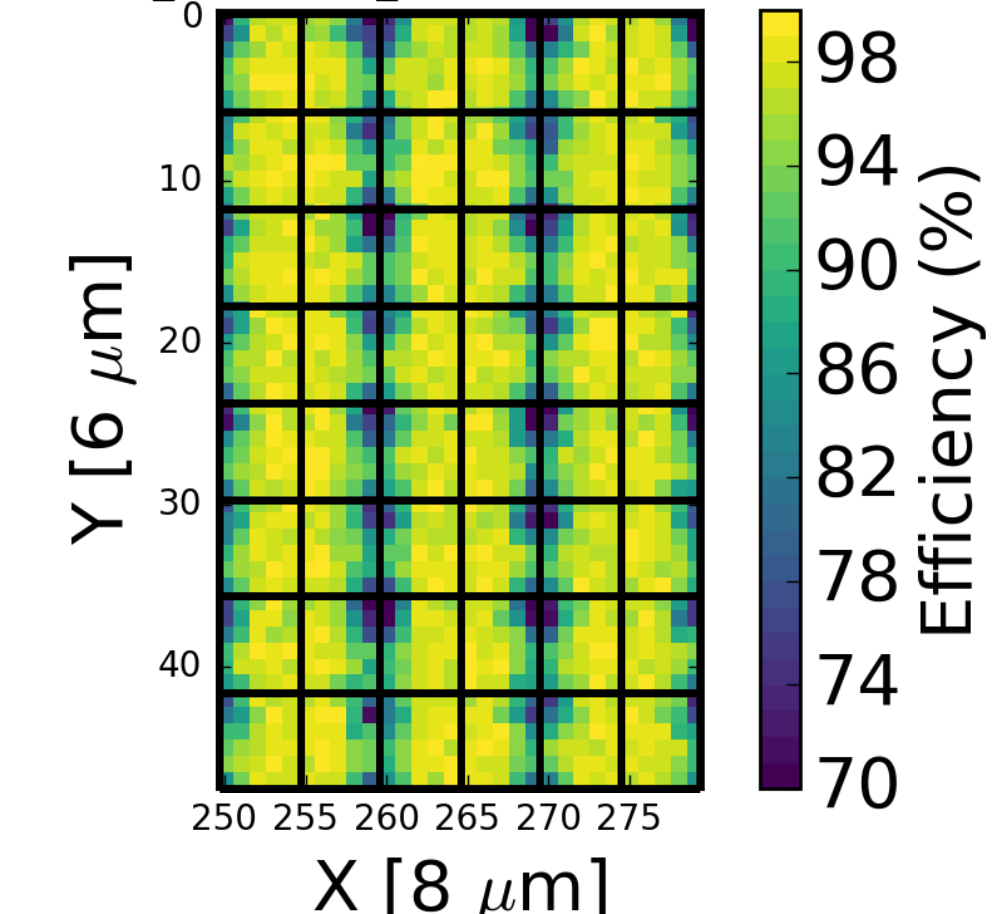}&
		\includegraphics[align=c,width=0.23\linewidth]{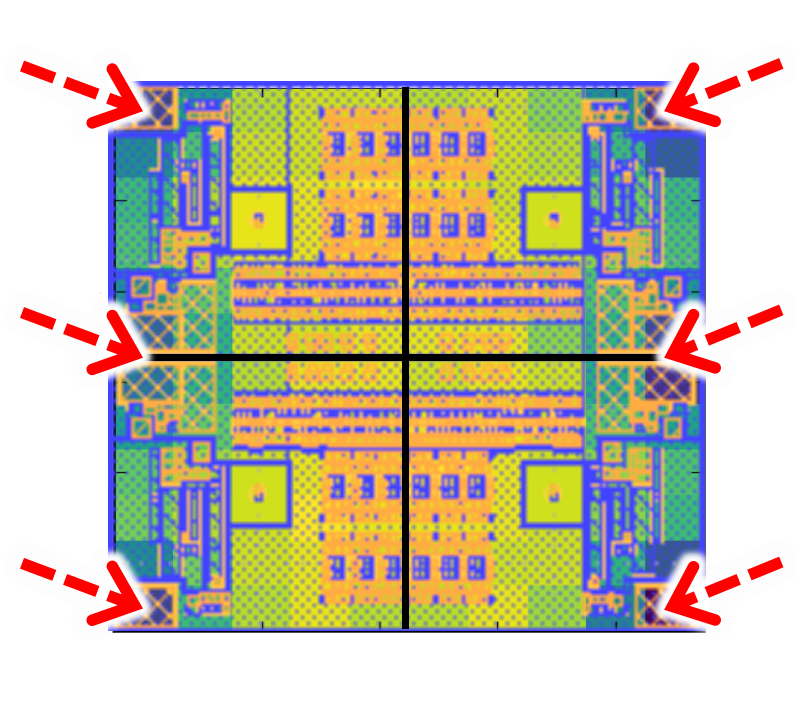}
		\\&\begin{tabular}{@{}c} \textbf{(a)} Efficiency maps in an\\ 8$\times$6 pixel array \end{tabular}& \begin{tabular}{@{}c} \textbf{(b)} Correlation to active\\regions in a 2$\times$2 pixel layout \end{tabular}\\
	\end{tabular}
	\caption{\label{fig:tjinpixefficiency} In-pixel efficiencies and correlation of corner losses to dense active regions in TJ-Monopix. Top: Pixels with removed deep p-well coverage. Bottom: Pixels with full deep p-well coverage.}
\end{figure}

A sub-pixel look into the hit detection efficiency before irradiation showed that the losses in TJ-Monopix were mainly localized at pixel corners of every double column (As observed in Figure ~\ref{fig:tjinpixefficiency}(a)). By looking at the pixel layout, it was possible to correlate these inefficient regions to dense active areas where large decoupling capacitors were placed (As depicted in Figure~\ref{fig:tjinpixefficiency}(b)). Moreover, the comparison between efficiency maps of the RDPW and FDPW sections showed that the inefficiencies at the pixel corners appeared only in one out of every two pixel rows for the RDPW layout. Based on these observations, it is suspected that the large active areas affect the local doping and reduce the -already small- lateral field and charge collection in the corners. The difference according to pixel layout suggests that the additional lateral field and charge generated by the removal of coverage compensated the field degradation near the edges. 

A correlation between deep p-well coverage and efficiency was also reported for a chip with similar front-end, pixel size and layout in the same CMOS process.~\footnote{R. Cardella et al., "MALTA: an asynchronous readout CMOS monolithic pixel detector for the ATLAS High-Luminosity upgrade" at PIXEL2018.} Furthermore, complementary simulations suggest modifications at the pixel edges that might substantially improve charge collection, which are currently prototyped.~\footnote{M. Munker et al., "Simulations of CMOS sensors with a small collection electrode improved for a faster charge-collection and increased radiation tolerance" at PIXEL2018.}



\section{Conclusions and Outlook}

A fast synchronous read-out architecture was successfully implemented in two fully monolithic depleted active pixel sensors in different CMOS processes and large or small electrode design approaches. Both chips were functional after wafer thinning down to $\mathrm{100\,\mu}$m and backside processing in the case of LF-Monopix. Moreover, they also remained operational -when cooled down- after irradiation with neutrons up to a dose of $1\times10^{15} \mathrm{\,n_{eq} / cm}^{2}$. Representative results reported for both chips before and after irradiation are summarized in Table~\ref{tab:monosummary}.

\begin{table}[htbp]
\small
\centering
\begin{tabular}{|c|c|c|c|c|}
\hline
	& \multicolumn{2}{|c|}{\textbf{LF-Monopix01}} & \multicolumn{2}{|c|}{\textbf{TJ-Monopix01}} \\
\hline
\textbf{DMAPS type}	& \multicolumn{2}{|c|}{Large electrode design} & \multicolumn{2}{|c|}{Small electrode design} \\
	& \multicolumn{2}{|c|}{(150nm CMOS / LFoundry)} & \multicolumn{2}{|c|}{(180nm CMOS, mod. / Towerjazz)} \\
\hline
\textbf{Dimensions}	& \multicolumn{2}{|c|}{$\mathrm{1\times1\, cm^{2}}$} & \multicolumn{2}{|c|}{$\mathrm{2\times1\, cm^{2}}$} \\
\hline
\textbf{Pixel size}	& \multicolumn{2}{|c|}{$\mathrm{250\times50\,\mu m^{2}}$} & \multicolumn{2}{|c|}{$\mathrm{40\times36\,\mu m^{2}}$} \\
\hline
	&     \textbf{Non-Irradiated}      &    \boldmath$1\times10^{15} \mathrm{\,n_{eq} / cm}^{2}$      &     \textbf{Non-Irradiated}      &    \boldmath$1\times10^{15} \mathrm{\,n_{eq} / cm}^{2}$                \\
\hline
\textbf{Signal MPV} &     $\mathrm{\sim23.3}$ ke-  $\mathrm{(@130\,V)}$      &    $\mathrm{\sim4.6}$ ke-  $\mathrm{(@130\,V)}$      &     $\mathrm{\sim 1.6}$ ke-      &     $\mathrm{\sim 1.4}$ ke-     \\
\hline
\textbf{ENC} &      $\sim200\pm50$ e-     &     $\sim350\pm50$ e-      &     $\sim16\pm2$ e-       &     $\sim23\pm3$ e-      \\
\hline
\textbf{Min. Threshold} &     $1400\pm100$ e-      &     $1700\pm130$ e-     &     $348\pm33$ e-      &      $569\pm66$ e-   \\
\hline
\textbf{Mean effciency} &     $99.6\%$      &     $98.9\%$     &     $97.1\%$      &      $69.4\%$   \\
\hline   
\end{tabular}
\caption{\label{tab:monosummary} Summary of representative features measured in LF-Monopix and TJ-Monopix.}
\end{table}

The results from measurements with LF-Monopix were overall promising and within expected values for a large electrode design, even after NIEL damage up to $1\times10^{15} \mathrm{n_{eq} / cm}^{2}$. At a matrix level, its performance is already comparable to the FE-I3 chip, though precise timing studies are still under way for a complete picture. In parallel, there are ongoing studies that expect to optimize the pixel layout and reduce the unit size in future submissions. By doing so, new prototypes would benefit from the timing and noise improvements due to a smaller detector capacitance. Further feedback from the performance of every front-end implementation after TID damage is still needed to determine the best performing amplifier and discriminator implementations.

In the case of TJ-Monopix, the non-irradiated chip was fully operational and efficient with a clearly discernible signal. After NIEL damage, the chip remained functional but the degradation in noise and localized inefficiencies suggest that the current front-end and pixel design require further optimization. The noise, minimum operational threshold and their dispersion should be reduced and tuned to enhance the signal-to-noise ratio after irradiation. In addition, a careful placement of active components and reduced deep p-well coverage would enhance charge collection. There are ongoing tests of proposed modifications in a small prototype, in order to evaluate whether additional fixes to the TJ modified process would improve the horizontal field near the pixel edges.

\acknowledgments

This research project was supported by the European Union's Horizon 2020 Research and Innovation programme under grant agreements No. 654168 (AIDA-2020) and 675587
(Marie Sk\l{}odowska-Curie ITN STREAM), and the German BMBF under grant No. 05H15PDCA9.


\end{document}